\documentclass[pre,showpacs,twocolumn,superscriptaddress]{revtex4}
\usepackage{amsmath,amsfonts,amssymb,wasysym}
\usepackage{bm}
\usepackage[dvipdfm]{graphicx}
\usepackage{color}

\newcommand{\cF}{ {\cal F} }

\begin{document}
\title{
DNA as a one-dimensional chiral material: Application to the 
structural transition between B form and Z form
}
\author{Teruaki Okushima}
\email{okushima@ike-dyn.ritsumei.ac.jp}
\affiliation{Department of Physics, Ritsumeikan University, 
Kusatsu City, Shiga 525--8577, Japan}
\author{Hiroshi Kuratsuji}
\affiliation{Research Organization of Science and Engineering, 
Ritsumeikan University, Kusatsu City, Shiga 525--8577, Japan}

\date{\today}
\begin{abstract} 
A dynamical model is presented for chiral change in DNA molecules.  
The model is an extension of the conventional elastic model which
incorporates the structure of base pairs and uses a spinor representation for the DNA configuration 
together with a gauge principle.  
Motivated by a recent experiment 
reporting chiral transitions between right-handed B-DNA and left-handed
 Z-DNA
[M. Lee, et.~al., Proc. Natl. Acad. Sci. (USA) {\bf 107}, 4985 (2010)],
we analyze the free energy for the particular case of linear DNA with an
 externally applied torque.
The model shows that there exists, at low temperature,
 a rapid structural change depending on the torque exerted on the DNA, 
which causes switching in B and Z domain sizes. 
This can explain the frequent switches of DNA extension observed in experiments.
\end{abstract}

\maketitle

\section{Introduction}
Recent advances in experimental techniques have made it possible to
perform experiments on the 
mechanical response of DNA molecules involving stretching and twisting of single DNA 
molecules \cite{recentReviewStrick,recentReviewMarko}. 
It has been observed that external forces can bring about changes in the macroscopic 
conformation as well as the base-pair structure of DNA. 
For example, frequent structural changes have been observed between right-handed 
B-DNA and left-handed Z-DNA when minute negative torque is exerted on a DNA molecule 
\cite{BZtrans}.
 In the regime of larger force, other structures, such as S-DNA, P-DNA, etc., are 
found \cite{tenYears}, and the coexistence of these structures in stretch-twist diagrams 
has also been reported \cite{structureTransition}.

Along with the experimental studies, various theoretical attempts have been made 
to describe the mechanical responses of DNA. 
One typical theory known as the elastic rod model enables a description in terms of 
a simple mechanism of supercoiling and its effects 
\cite{recentReviewMarko,MarkoSiggia,Strick,rodPlectoneme}. 
Other significant theoretical models, such as the Poland--Scheraga
model \cite{PS} and 
the Peyrard--Bishop--Dauxois model \cite{PB}, are concerned with the problem of 
denaturation, and have been extended so as to include the coupling between DNA twisting 
and denaturation \cite{cocco1999}.
The interaction between DNA supercoiling and denaturation has been studied by 
extending the classical elastic rod model \cite{supercoilDenaturation}.
An effective potential model, which predicts twist-stretch coupling in
the mechanical 
response of B-DNA, was presented in \cite{twistStretch}.
Statistical models that phenomenologically describe various structural transitions 
were developed in \cite{statModel}. 
Several mesoscopic models have also been developed, which can describe the interaction between DNA
conformation  and {\em melting} structural transition \cite{yan_marko,palmeri}.
However, 
there has not been a mechanical model of DNA that describes 
the interplay between the global configuration and the various intrinsic structures 
of DNA base-pairs, including B, Z, S structures.
\begin{figure}[t]
\includegraphics[width=5.5cm]{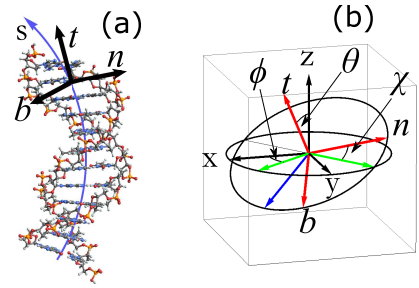}
\caption{(Color online) 
(a)	The DNA molecule consists of two sugar-phosphate chains bridged by base pairs, 
forming the double helix. Moving frame vectors $({\bm b,\bm n,\bm t})$ are defined 
along the helical axes parameterized by a continuous parameter $s$.
(b) The frame vectors are specified by the depicted Euler angles
 $ (\theta, \phi, \chi) $.
}
\label{manga}
\end{figure}

The purpose of this article is to propose a phenomenological model that combines 
the aspects of conformation and intrinsic structure to obtain a clear picture of DNA 
mechanics. 
Our model adopts the Landau theory, which consists of the free energies for helical 
structures of base pairs connected by hydrogen-bonds, elastic deformations of DNA 
configuration, and the interaction between them. 
The interaction is introduced by using the gauge principle,
with the 
spinor representation of DNA configuration.
Monte Carlo simulation of the model is used to examine B-Z structural transitions 
of linear DNA under the application of external torques. 
It is found that there exists a cooperative effect, depending on both the temperature 
and the torque exerted on the DNA, which can be described in terms of the probability 
to create kink-antikink pairs. This cooperative effect causes sharp switching in B 
and Z domain sizes, which can explain the frequent switches of DNA extension observed 
in recent experiments \cite{BZtrans}.

\section{Model } 
We use two kinds of order parameters: 
one is the order parameter for the conformations of DNA and the other is that for 
the intrinsic base-pair structures, which are determined by the hydrogen bonds between 
them. 
We first construct a moving frame that describes the conformation of the 
double-stranded DNA of $N$ base-pairs (bp's). 
Figure \ref{manga}(a) shows a continuous parameter $s$ representing the number of bases 
from one end of the molecule, where the DNA helical axis is parametrized as ${\bm 
r}(s)$. 
The normalized tangent vector along the axis is defined as 
$ {\bm t } = \frac{1}{v} \frac{d {\bm r}}{ds}$,
where $v$ is the increase per bp along axis $\|d {\bm r}/ds\| $.  
A moving frame is defined at each point $s$, by the set of three mutually orthogonal 
vectors 
$({\bm b}, {\bm n},{\bm t})$,
where ${\bm b}$ is the normalized vector in the direction of base position 
and $\bm n$ is specified as $ {\bm n} = {\bm t}\times {\bm b}$ to form a right handed 
coordinate system.

We now introduce an order parameter to describe the configuration of the moving frame 
$({\bm b}, {\bm n},{\bm t})$, 
which is represented by a spinor $ \Psi $:
$\Psi=(\psi_1,\psi_2)\in \mathbb{C}^2$ 
with $|\psi_1|^2+|\psi_2|^2=1$.
The spinor is parameterized by a set of Euler angles, as in \cite{LL}.
Using Euler angles $ (\theta, \phi, \chi) $ as depicted in Fig.~\ref{manga}(b),
the spinor is written as 
\begin{equation}
\Psi=(
\cos\frac{\theta}{2}e^{-i\frac{\phi+ \chi}{2}},
\sin\frac{\theta}{2}e^{i\frac{\phi- \chi}{2}})^t 
\end{equation}
and 
$ {\bm t} $ can be written as an average of the Pauli
vector $ {\bm \sigma} $ by the spinor: 
$
\bm{t}=
\Psi^{\dagger}\bm{\sigma} \Psi  = 
(\sin \theta \cos \phi ,\sin \theta \sin \phi , \cos \theta )
$,
where ${\bm \sigma}$ is given by  $(\sigma^1,\sigma^2,\sigma^3)$
with the following Pauli matrices:
\begin{equation*}
\sigma^1=
{
\begin{pmatrix}0&1\\1&0\end{pmatrix}
},\ 
\sigma^2=
{
\begin{pmatrix}0&-i\\i&0\end{pmatrix}
},\ 
\sigma^3=
{
\begin{pmatrix}1&0\\0&-1\end{pmatrix}
}
.
\end{equation*}

We next consider an explicit form for the free energy of the DNA conformation,
in terms of a functional of $\Psi$.
To this end,
we set the following requirements
(i) rotational symmetry and
(ii) inclusion of up to first derivative with respect to $s$.
Then, 
the simplest form satisfying these criteria is given by 
$F= \int_0^N  \cF  ds$,
where
\begin{equation}
\cF=\cF_s \equiv
\frac{k_1}{2} \frac{d\Psi^{\dagger}}{ds}\frac{d\Psi}{ds} 
+   \frac{k_2}{2}
 \left|  \Psi^{\dagger}\frac{d\Psi}{ds} \right|^2. 
\label{Lrod}
\end{equation}
With the use of the Euler angles,
$\cF_s$ 
reads
$
   \frac{B}{2}[(\frac{d\theta}{ds})^2 +(\frac{d \phi}{ds})^2 \sin^2\theta]
   +\frac{C}{2} (\frac{d \phi}{ds}\cos \theta +\frac{d \chi}{ds} ) ^2,
$
with $B=k_1/4,\ C=(k_1+k_2)/4$,
which is the well-known free-energy density of an elastic rod with isotropic bending 
elasticity $B$ and torsional elasticity $C$ \cite{LLelastic}.
Note here that, if an external force $f$ in the  $z$-direction and torque $\tau$ are 
applied to the DNA ends, the free energy density becomes
$ \cF_s-f v \cos (\theta) -\tau L_k$,
where $L_k$ is the linking number of the DNA \cite{MarkoSiggia}.

Now we introduce the order parameter $\rho$ representing the internal structures of 
base pairs, 
 via gauge coupling with the conformational 
spinor. 
Consider the gauge transformation $ \Psi \to \Psi \exp(i \alpha) $.  
If $ \alpha $ is constant, the invariance of the free energy under the transformation 
is apparent, but if $ \alpha $ has $s$-dependence, the local gauge invariance does 
not hold. 
To keep the gauge invariance \cite{deGennes}, 
we define the order parameter $\rho$ for the internal structure of base pairs as a 
gauge field, which we call {\em chiral field}.
$\rho$ is the rotational angle between successive base pairs in the helix, 
$\rho>0$ for right handed rotation and $ \rho<0 $ for left handed rotation. 
After progressing $ ds $, we have $\chi \to \chi+ \rho ds$ and 
\begin{equation*}
\Psi^{\parallel}(s+ds) 
= 
(\psi_1 e^{-i  \frac{\rho}{2} ds},\psi_2 e^{-i  \frac{\rho}{2} ds})^t
= 
\Psi-i \frac{\rho}{2}  \Psi ds,
\end{equation*}
which just gives the parallel transport. 
Since the displacements of sugar-phosphate strings 
from the parallel transport $\Psi^{\parallel}(s)$
gives rise to an elastic potential,
we make the replacement of $d/ds \to D/Ds$ in Eq.~(\ref{Lrod}),
where
$D/Ds$ is the covariant derivative such that
\begin{equation*}
\frac{D \Psi}{Ds} \equiv 
\lim_{ds \to 0} \frac{\Psi(s+ds)-\Psi^{\parallel}(s+ds)}{ds}
=\frac{d \Psi}{ds} +i \frac{\rho}{2} \Psi
.
\end{equation*}
The resulting free-energy density becomes 
$\cF_s+\cF_{s\rho}$, 
where
$ \cF_{s\rho}= -iC \rho(\Psi^{\dagger}\frac{d\Psi}{ds} - 
\frac{d\Psi^\dagger}{ds} 
\Psi
)+C\frac{\rho^2}{2}
$; 
thereby
we have a new coupling term $\cF_{s\rho}$ between
the spinor field and the chiral field.
In terms of the Euler angles, the coupling term is given by 
\begin{equation*}
\cF_{s\rho}=-C\rho( \frac{d\chi}{ds}+ \frac{d\phi}{ds} \cos 
\theta)+\frac{C\rho^2}{2},
\end{equation*}
from which we see that $C\rho $ is conjugate to 
the angular velocity with respect to $s$ around ${\bm t}$
($ \Omega_3 \equiv \frac{d\chi}{ds}+ \frac{d\phi}{ds}\cos \theta $). 
Combining the above two terms, the total free-energy density $
\cF_{s+\rho} $  can be written in terms of the Euler angles:
 \begin{equation}
  \cF_{s+\rho}  =    
   \frac{B}{2}[(\frac{d\theta}{ds})^2 +(\frac{d\phi}{ds})^2
   \sin^2\theta)]
   +\frac{C}{2}( \frac{d\phi}{ds}\cos \theta +\frac{d\chi}{ds}  - \rho
		) ^2.
\label{gaugeRod}
\end{equation} 
This is similar to the elastic rod model
\cite{MarkoSiggia},
but differs in the point that there is a coupling 
with the chiral field $ \rho $.
The conventional elastic rod model
does not allow the inclusion of such a coupling term,
because it does not take into account the
deformation degrees of freedom coming from the base pairs, 
whereas in our case the spinor representation 
enables us to take into account such degrees of freedom
naturally as a gauge field through the covariant derivative. 

Now Landau's free energy for base-pair structures 
is defined by a general functional 
of the above-introduced chiral field $\rho(s)$
and the length scale $v(s)=\|d{\bm r}/ds\|$. 
The free-energy density $\cF_b$ is given by
\begin{equation}
\cF_{b} = 
\frac{d_1}{2}\big(\frac{d\rho}{ds}\big)^2
+\frac{d_2}{2}\big(\frac{dv}{ds}\big)^2+\dots
+V(\rho,v,\dots).
\label{general}
\end{equation}
Here the last term is the effective potential,
whose local minima,
$(\rho_\text{B},v_\text{B},\dots)$,
$(\rho_\text{Z},v_\text{Z},\dots)$,
$(\rho_\text{S},v_\text{S},\dots)$, etc.,
 can be assigned to 
the locations of the base-pair structures B-DNA, Z-DNA, S-DNA, etc.,
respectively.
On the other hand,
the gradient terms represent additional free-energy costs 
due to changing base-pair structures along $s$.

In this way, we have constructed the total free-energy density for DNA
as 
\begin{equation}
\cF=\cF_{s+\rho}+\cF_b.
\end{equation}
It should be noted that
the coefficients $B, C$ can be extended 
to be functions of the order parameters for base-pair structure \cite{endnote}.
Our model contains the conventional elastic model,
as a limiting case of  the rigid base-pair limit, 
  $V(\rho) = -\delta (\rho-\omega_0)$.
Namely,
the latter is obtained by
substituting $\omega_0$ for $\rho$ in
Eq.~(\ref{gaugeRod}).
In the rigid sugar-phosphate limit $C \to \infty$,
it recovers the model given in Ref.~\cite{phi4},
becoming nonlocal in the sense that
it contains the second derivative of the order parameter.

\section{Chiral transition induced by external torque } 
\begin{figure}[b]
\includegraphics[width=5.2cm]{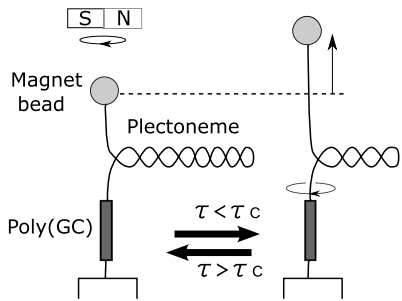}\\
\caption{
Schematic representation of the experiment in Ref.~\cite{BZtrans}
(see text). 
}
\label{experiment}
\end{figure}

As the first application of our model to realistic conditions,
we consider the experiment performed by Lee, et.al., in  Ref.~\cite{BZtrans}.
In the experiment, shown schematically in Fig.~\ref{experiment}, 
the ends of a DNA molecule were fixed with 
a glass cover slip and a magnetic bead, respectively.
By applying negative torque to the DNA with the bead,
minute negative superhelicity
was induced in the DNA
and then
an inter-winding DNA configuration, called {\em plectoneme}, was formed.
Below the plectoneme, in the linear part of the DNA, there was
a poly(GC) sequence that tends to change structures between B-DNA and Z-DNA.
They measured frequent sharp changes of 
the DNA extension between the cover slip and the bead, as well as
frequent B-Z transitions in the poly(GC) part.
Therein,
the B-Z transitions induce winding and unwinding of the plectonemic
conformation,
resulting in the observed sharp switches between small and large DNA extensions, 
respectively.

To analyze this experiment,
it is sufficient to consider a model of
a linear DNA molecule with B-Z structural transitions.
By setting  $\theta=0,\phi=0$ in Eq.~(\ref{gaugeRod}),
we obtain the total free-energy density given by
\begin{equation}
\cF=\frac{C}{2}\left(\frac{d\chi}{ds}- \rho \right)^2+
\frac{d_1}{2}\big(\frac{d\rho}{ds}\big)^2
+V(\rho)-\tau  \frac{d\chi}{ds},
\end{equation}
where we adopt the following values:
$C=91 \times 10^{-20}$J,
$d_1=D_1/\omega_0^2$ with $D_1=4.1 \times 10^{-21}$ J and 
$\omega_0=0.6$ rad/bp,
and
$\tau$ is the torque applied to the end of the linear DNA molecule.
The potential density $V(\rho)$
is given by
\begin{equation}
V(\rho) = V_0[(\rho/\omega_0)^2 - 1]^2+\tau_c {\rho},
\end{equation}
with
$\tau_c=-7.9 \times 10^{-21}$J \cite{marko2007} 
and
$V_0 = 6  \times 10^{-20}$J,
where the B-DNA and Z-DNA conformations
correspond to the minima at around
$\rho=\omega_0$ and
$\rho=-\omega_0$, respectively.
These parameter values are determined as follows.
We expect the twist persistence length is 75nm,
which gives
$
C=75 \text{nm}\times k_B T_R/0.34\text{nm}
=221 k_B T_R =91 \times 10^{-20}\text{J}
$,
where 
$T_R$ denotes room temperature and
$k_B T_R=1.38\times 10^{-23}\text{JK$^{-1}$}\times300
\text{K}=4.1\times 10^{-21}$J is used.
In our model,
$\frac{d_1}{2}(d\rho/ds)^2$ gives
the energy cost for structure change,
which was already given in Ref.~\cite{phi4} by $2 k_B T_R=8.2\times 10^{-21} $J.
From the equality
$\frac{d_1}{2}(2 \omega_0)^2=2 k_B T_R$,
one obtains
$d_1=D_1/\omega_0^2$,
where $D_1=k_B T_R=4.1\times10^{-21}$J.
As to $V_0$,
we adopt the value
$V_0=6\times 10^{-20} \text{J}(\sim 15 k_B T_R)$ 
according to  Ref.~\cite{phi4}.
With this value,
the domain wall energy of our model 
($E_{\text{B-Z}}=8/3\sqrt{2 d_1 V_0 \omega_0}$ \cite{kinknote0})
agrees with the ``B-Z junction energy''  5 Kcal/mol, 
that was estimated in Ref.~\cite{bzjunction}.
The free energy of Z conformation is $E_\text{Z}=2.3 k_B T_R$ \cite{statModel},
thereby
$\tau_c$ satisfies $2 \omega_0 \tau_c  = -2.3k_B T_R$,
which gives
$\tau_c=-7.9 \times 10^{-21}$.
Note that we have chosen the depths of minima
that correspond to normal physiological conditions
\cite{saltNote}. 
Furthermore, for simplicity,
we have neglected the
$v$-dependency supposed in the general expression of Eq.~(\ref{general}). 

\begin{figure}[t]
\includegraphics[width=6.6cm]{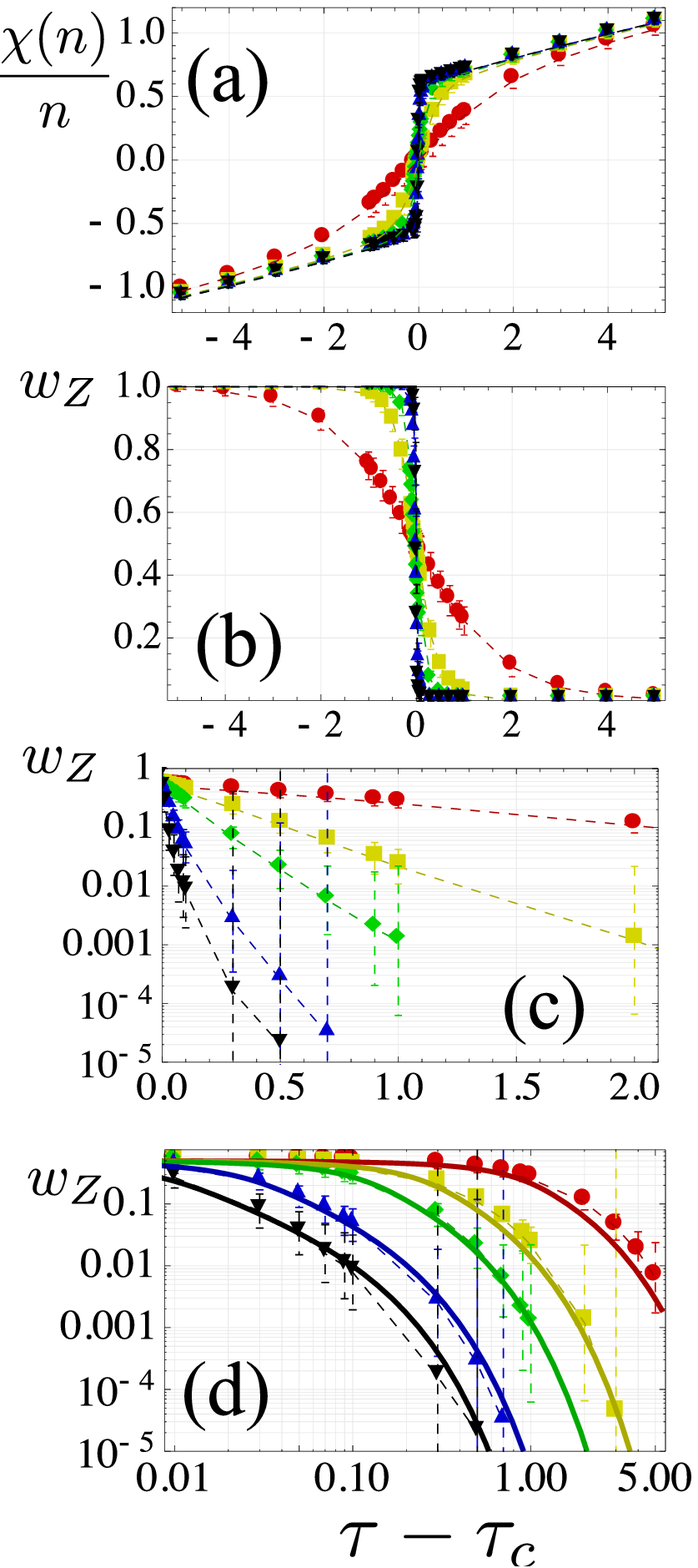}\\
\caption{(Color online) 
Exchange Monte Carlo results
for $N=100$ with 1000 samples:
(a)$\chi_n/n$ ($n=50$),
(b)$w_\text{Z}$, 
(c)Log plot of $w_\text{Z}$, 
(d)Log-Log plot of $w_\text{Z}$, 
as functions of $\tau-\tau_c$
for $\beta
 =1$($\CIRCLE$, red), 3($\blacksquare$, yellow),
5($\blacklozenge$, green), 10($\blacktriangle$,
 blue), and 14($\blacktriangledown$, black) 
in units of $V_0$($=6\times 10^{-20}$J),
where
the vertical error bars on data points represent their variances.
$w_\text{Z}$ is symmetric with respect to $(\tau,w_\text{Z})=(\tau_c,0.5)$.
The $i$-th bp is considered to be Z-structure if $\rho_i<0$.
Solid lines in (d) are theoretical curves of Eq.(\ref{th_wz})
with fitted values 
$E_{\text{B-Z}}=0.11 V_0$ and
$\gamma=0.38$ 
\cite{kinknote0}.
Between $\chi(n)$ and $w_\text{Z}$, the approximate relation
$
d\chi/ds \simeq \rho_\text{B}+(\rho_\text{Z}-\rho_\text{B}) w_\text{Z}+\tau/C
$
holds,
where $\rho_\text{B}=0.6$, $\rho_\text{Z}=-0.6$.
}
\label{slope_t}
\end{figure}
To perform numerical simulation,
we discretize the parameter $s$.
The resulting free energy is given by
\begin{eqnarray*}
F&=&
\sum_{i=1}^{N}
[
V(\rho_i)
+\frac{C}{2}(\chi_{i}-\chi_{i-1}-\rho_i )^2\\
&&
+
\frac{d_1}{2}(\rho_{i}-\rho_{i-1})^2
-\tau (\chi_{i}-\chi_{i-1})],
\end{eqnarray*}
with $\chi_0=0$ and $\rho_0=\omega_0$.
By using the replica exchange Monte Carlo method \cite{exchangeMC},
we sampled the equilibrium states,
under various conditions of
the external torque $\tau $ and  temperature $T$.
With every step, this method 
updates all replicas $X_r$ 
with different temperatures $T_r $ ($r=1,2,\dots$) simultaneously
and then exchanges these replicas, 
to enhance the ergodicity of these samples.
In the simultaneous updates,
each replica
$X_r=(\rho^{(r)}_1,\Delta\chi^{(r)}_1,\rho^{(r)}_2,\Delta\chi^{(r)}_2,\dots,\rho^{(r)}_N,\Delta\chi^{(r)}_N)$,
where $\Delta \chi^{(r)}_i \equiv \chi^{(r)}_i-\chi^{(r)}_{i-1}$
(i.e.,
$\chi^{(r)}_i=\sum_{j=1}^i \Delta \chi^{(r)}_j$)
follows  the following steps:
\begin{description}
\item[Step~1:]
Draw a random integer $i$ from 1 to $N$.
\item[Step~2:]
$\Delta \chi^{(r)}_i$ and $\rho^{(r)}_i$
are supposed to be shifted to
$ \tilde{\rho}^{(r)}_i =\rho^{(r)}_i+\epsilon_\rho \ R_{-1,1}$ 
and
$\Delta\tilde{\chi}^{(r)}_i=\Delta \chi^{(r)}_i+\epsilon_\chi \ R_{-1,1}$,
where 
$\epsilon_\rho$ and 
$\epsilon_\chi$ are 
the values of maximum shifts for $\rho$ and $\Delta \chi$, respectively,
and
$ R_{-1,1}$ is a random number between $-1$ and $1$.
\item[Step~3:]
Compute the free-energy change $\delta F =F(\tilde{X}_r)-F(X_r)$
and 
$W=\exp(-\beta_r \delta F)$,
where $\beta_r=(k_B T_r)^{-1}$ with Boltzmann's constant $k_B$.
\item[Step~4:]
If $W>R_{0,1}$ then
make changes
$\Delta \chi^{(r)}_i=\Delta\tilde{\chi}^{(r)}_i$ and 
$\rho^{(r)}_i=\tilde{\rho}^{(r)}_i$;
otherwise do not change.
\end{description}
After these simultaneous updates,
for a randomly chosen neighboring pair of replicas, say  $X_r$ and $X_{r+1}$, 
exchange their configurations 
by the following acceptance criterion:
\begin{description}
\item[Step~1:]
Compute the cost function 
\begin{equation*}
\Delta =(\beta_{r+1}-\beta_{r})[F(X_r)-F(X_{r+1})]
\end{equation*}
and 
$W=\exp(-\Delta)$.
\item[Step~2:]
If $W>R_{0,1}$ then
exchange replicas as
$(X_r,X_{r+1}) =(X_{r+1},X_r)$; otherwise do not exchange these replicas. 
\end{description}
In the simulations,
we set
$\{\beta_r\}=$
$\{$0.01, 0.03, 0.05, 0.07, 0.1, 0.3, 0.5, 0.7, 1, 1.25, 1.5, 1.75, 2, 
2.5, 3, 3.5, 4, 4.5, 5, 5.5, 6, 6.5, 7, 7.5, 8, 8.5, 9, 9.5, 10, 
10.5, 11, 11.5, 12, 12.5, 13, 13.5, 14$\}$,
and 
confirmed the frequent exchanges of replicas.
All the states are sampled at every $100N$ exchange Monte Carlo  steps,
after equilibration.
\begin{figure}[t]
\begin{center}
\includegraphics[width=6cm]{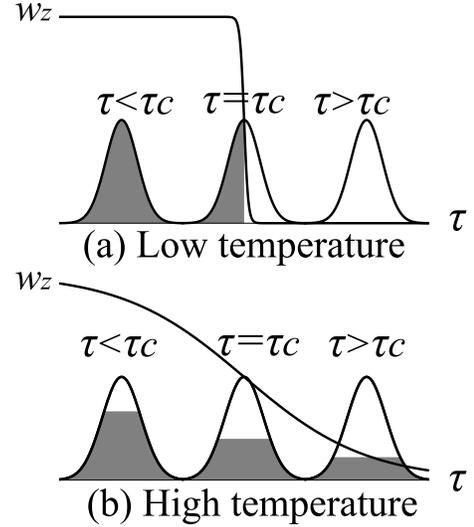}
\end{center}
\caption{
Schematic illustration of 
torque responses of $w_\text{Z}$.
(a) At low temperature,
$w_\text{Z}$ behaves like switch.
Three curves represent there cases of torque  distributions:
the average $\tau$ satisfies
$\tau<\tau_c$
(left),
$ \tau=\tau_c$
(middle),
and 
$ \tau >\tau_c$
(right).
The fluctuations $\delta \tau$ are given by Eq.~(\ref{delta_tau}).
Shaded regions in the distributions
represent the ratios of Z-DNA conformations.
(b)
At high temperature,
$w_\text{Z}$ changes gradually as a function of $\tau$,
compared to $\delta \tau$.
Hence,
the ratios of Z conformations also change gradually,
and do not show  switching behaviors.
}
\label{manga2}
\end{figure}

Fig.~\ref{slope_t}(a)
shows the average twists as functions of $\tau$.
We see that,
at high temperatures compared to $V_0$,
the response to $\tau$
is  approximately uniform.
In contrast, at relatively low temperatures,
$\chi_N$ are susceptible to small differences in $\tau$
especially at around $\tau=\tau_c$.
This result is in accord with the experiment of Ref.~\cite{BZtrans}.
Namely,
in the experiment,
the torques applied to the linear part of DNA
can not be controlled to be a certain value
but actually fluctuate around the value as
\begin{equation}
\delta \tau\sim \sqrt{C k_B T/N} \sim 0.032 V_0,
\label{delta_tau}
\end{equation}
due to
the disturbance from
the rest of the DNA and/or
the physiological buffer,
and thus
$\chi_N$ changes widely
due to the high susceptibility to $\tau$ at around $\tau_c$,
since the experiment was carried out at low temperatures ($\beta V_0 \sim 14$).
The large changes in $\chi_N$ alternate
the winding number of the plectoneme,
resulting in the change of DNA extension observed in the experiment. 

We shall now confirm that
the sensitivity of $\chi_N$ to $\tau$ is caused by the B-Z transition.
For this, the Z-DNA ratio, $w_\text{Z}$, is plotted 
as functions of $\tau$ in Fig.~\ref{slope_t}(b)-(d).
This clearly
shows that
the sharp sensitivity of $\chi_N$ to $\tau$ at low temperatures 
found in Fig.~\ref{slope_t}(a) is coincident with
the sharp decrease of $w_\text{Z}$,
while
the mild dependency of $\chi_N$ on $\tau$ at high temperatures 
is accompanied with slow decrease of $w_\text{Z}$.
Hence,
our model clearly shows that
the experimentally found sharp changes can be attributed to
the dependency of B-Z transitions on torque,  
which becomes singular in the low temperature limit.

Let us discuss 
the details of
the structural transition induced by external torque
and the resulting singular responses, shown in Fig.~\ref{slope_t}.
In low temperature such that $T/V_0 \sim 0$ 
(room temperature satisfies this condition), 
the response to torque  is sharp, and then behaves like a switch.
Hence, for example, 
if $\tau-\tau_c>0$ is suddenly changed to $\tau-\tau_c<0$, 
then
all DNA structures nucleate from false ground state (B-DNA) 
to true ground state (Z-DNA).
This point is illustrated in Fig.~\ref{manga2}(a).
Otherwise,
if the superhelicity is set around ``the midpoint value,''
then the torque fluctuates at around $\tau \sim \tau_c$,
which induces
the rapid structural switching observed in Fig.~5A in Ref.~\cite{BZtrans}.
Namely, we have confirmed that 
this switching induces the changes both in $w_\text{Z}$ and in 
the change of 
DNA extension, $dL$.
In contrast, when the temperature is high ($T/V_0>1$),
the response becomes gradual, as shown in Fig.~\ref{manga2}(b).
In this case,
we can neither expect the {\em switching} behavior  in $w_\text{Z}$ nor $dL$.
In this way, with the use of our model,
we have succeeded in clarifying
the condition for stepwise change 
and for interconversion between these states
observed in Ref.~\cite{BZtrans}.

\section{Nucleation theory\label{nucl_sec}}
\begin{figure*}[t]
\includegraphics[width=15cm]{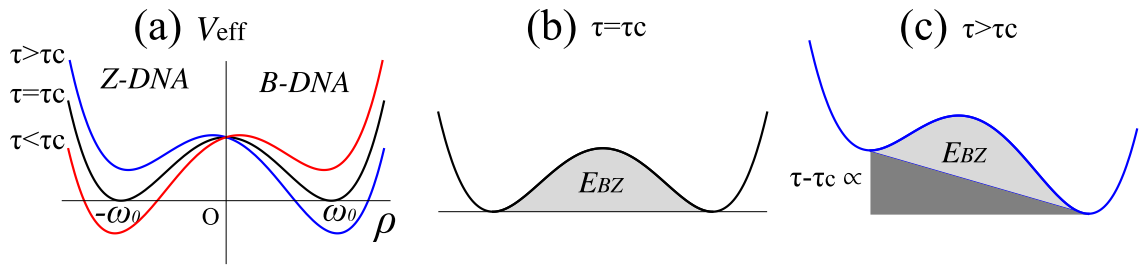}\\
\caption{(Color online) 
(a) 
Illustration for
$V_\text{eff}$ in Eq.~(\ref{veff_eq}) as functions of  $\rho$:
for $\tau<\tau_c$ (red),
$\tau =\tau_c$ (black),
and
$\tau >\tau_c$ (blue).
$V_\text{eff}$ is symmetric at $\tau=\tau_c$.
Figs.~(b) and (c) {\it schematically}  show 
the domain-wall energy given by Eq.~(\ref{p_eq}):
(b)
$E_{\text{B-Z}}$ is the domain wall energy 
for the symmetric $\tau=\tau_c$ case.
In this case, the domain-wall energy 
corresponds to the shaded region.
(c)
For $\tau \neq \tau_c$ case,
the domain wall energy requires the excess energy,
shown in darker shade of gray.
This figure shows that the excess energy is proportional to
 $|\tau-\tau_c|$,
as in Eq.~(\ref{p_eq})}
\label{veff}
\end{figure*}
Our findings can be explained in terms of
the probability of kink-antikink nucleation 
and the average size of the metastable domain \cite{Langer}.

The variational equation $\delta F=0$ 
reads
\begin{equation}
\frac{d \chi}{ds}-\rho-\frac{\tau}{C}=0
\end{equation}
and
\begin{equation}
d_1 \frac{d^2 \rho}{ds^2}=-\frac{d V_{\text{eff}}(\rho)}{d \rho},
\end{equation}
where
\begin{equation}
V_{\text{eff}}(\rho)=V_0\left[\left(\frac{\rho}{\omega_0}\right)^2-1\right]^2-(\tau-\tau_c)\rho.
\label{veff_eq}
\end{equation}
$V_\text{eff}$ is depicted in Fig.~\ref{veff}(a) for various $\tau$.
For $\tau>\tau_c$, B conformation is stable, while Z conformation is
unstable.
In contrast to this, for $\tau<\tau_c$, Z conformation is stable, while
B conformation is unstable.
At $\tau=\tau_c$,
$V_\text{eff}$ is symmetric and these conformations are equally probable.

The probability $P$ to form the nucleation is
given by
\begin{equation}
P=[1+e^{\beta E_{\text{nucl}}(\tau)}]^{-1},
\end{equation}
where 
$E_\text{nucl}(\tau)$ is the nucleation energy at torque $\tau$.
Since the nucleation starts with creating kink-antikink pairs,
$E_\text{nucl}$ is roughly estimated to be
twice the domain-wall energy  $E_\text{dom}$ \cite{kinknote}:
\begin{equation}E_{\text{nucl}}(\tau)=2 E_\text{dom}(\tau).
\label{Enucl}
\end{equation}
Then,
the torque-dependence of the domain wall energy is assumed
to be
\begin{equation}
E_{\text{dom}}(\tau)= E_{\text{B-Z}}+\gamma |\tau-\tau_c|
\label{p_eq}
\end{equation}
where 
$E_{\text{B-Z}}$ is the domain wall energy at $\tau=\tau_c$, shown in
Fig.~\ref{veff}(b),
and 
$\gamma$ is the coefficient of torque in the domain wall energy,
shown in 
Fig.~\ref{veff}(c).

The distance between centers of neighboring metastable domains
is estimated to be $1/P$.
Once a metastable domain is created, 
it grows to a size $r$ 
with the probability 
$\exp(-\beta \Delta V r)/{\cal N}$,
where
$\Delta V$ is the free-energy density difference
and
${\cal N}=\int_0^{1/P} \exp(-\beta \Delta V r) dr$.
The average domain size is estimated as 
\begin{eqnarray*}
\langle r \rangle 
&=&
\int_0^{1/P}
r e^{-\beta \Delta V r}/\cal{N} \\
&=&P^{-1}
\frac{1+x-\exp x}{x(1-\exp x)},
\end{eqnarray*}
with 
$x=\beta \Delta V/P$.
From these results,
the ratio of domain size between 
the metastable and the stable domain
is given by
$\langle r \rangle : P^{-1}-\langle r \rangle$.
If $\tau >\tau_c$,
then B-DNA is stable and Z-DNA is metastable
and thus
$
w_\text{Z}=
\langle r \rangle P
$.
Otherwise,
if $\tau <\tau_c$,
then Z-DNA is stable and B-DNA is metastable
and
$
w_\text{Z}=
1-\langle r \rangle P
$.
Putting these estimates together,
we obtain
\begin{equation}
w_\text{Z}=
\frac{1+x-e^x}{x(1-e^x)},
\label{th_wz}
\end{equation}
with 
\begin{equation}
x=\beta(V_\text{Z}-V_\text{B})/P,\quad
V_\text{Z}-V_\text{B}=2 \omega_0 (\tau-\tau_c).
\end{equation}
Let us confirm that Eq.(\ref{th_wz}) describes our numerical result of 
Fig.~\ref{slope_t}.
The solid curves in Fig.~\ref{slope_t}(d) plot this relation
(\ref{th_wz}),
which agree with the numerical calculations.  

We can also explain the singularity. 
Expanding $w_\text{Z}$  around $\tau = \tau_c$
gives
\begin{equation}
w_\text{Z} \simeq 0.5-P^{-1}\beta \omega_0 (\tau-\tau_c)/6,
\end{equation}
where the slope with respect to  $\tau$
is proportional to the average domain size $\sim P^{-1}$.
Hence,
the lower the temperature,
the smaller the kink-antikink creation rate.
Furthermore,
the larger the average metastable size,
the higher
the structural susceptibility to torque.
We therefore conclude that the temperature-dependent cooperative effect
is the origin of the observed singularity. 

This is analogous with the instability of a
one-dimensional (1D) magnetic system.
Actually,
we see the clear correspondence with the  1D ferromagnetic Ising model:
\begin{eqnarray*}
\tau-\tau_c &\leftrightarrow & \text{magnetic field},\\
\text{domain wall energy} &\leftrightarrow &\text{nearest neighbor
 interaction}. 
\end{eqnarray*}
Hence,
it is natural 
that
the role of cooperativity
(domain wall energy divided by thermal energy) is essential to
understand the results of Fig.~\ref{slope_t}.
Furthermore,
in the low temperature limit,
there occurs an instability at 
the external torque of $ \tau = \tau_c $, 
which, in the Ising system, corresponds to  zero external magnetic field
\cite{oneD}.

Finally,
we shall show 
that
the twist dependence of transition rate reported in the experiment of Ref.~\cite{BZtrans}
is qualitatively explained by this nucleation picture.
Based on the above arguments on the nucleation energy (\ref{Enucl}),
the nucleation energy from stable structure
and from metastable structure are, respectively, given by
\begin{eqnarray*}
E_\text{nucl}&=&2(E_{\text{B-Z}}+\gamma |\tau-\tau_c|),\\
E^\text{meta}_\text{nucl}&=&2(E_{\text{B-Z}}-\gamma |\tau-\tau_c|).
\end{eqnarray*} 
Hence,
the transition rates from B to Z structure and from Z to B structure, 
respectively,
turn out to be, for $\tau>\tau_c$,
\begin{equation}k_{\text{B-Z}}=t_\text{B}^{-1} e^{-\beta E_\text{nucl}},\quad
k_{\text{Z-B}}=t_\text{Z}^{-1} e^{-\beta E^\text{meta}_\text{nucl}};
\end{equation}
and, for  $\tau<\tau_c$,
\begin{equation}k_{\text{B-Z}}=t_\text{B}^{-1} e^{-\beta E^\text{meta}_\text{nucl}},\quad
k_{\text{Z-B}}=t_\text{Z}^{-1} e^{-\beta E^\text{meta}_\text{nucl}},
\end{equation}
where
$t_\text{B}, t_\text{Z}$ are the inverses of frequency factors for B and Z structures.
From these results,
the equilibrium constant $K_\text{eq}$ ($\equiv k_{\text{B-Z}}/k_{\text{Z-B}}$)
is given by
$\frac{t_\text{Z}}{t_\text{B}}e^{-4 \beta \gamma (\tau-\tau_c )}$.
By noting that
$\tau$ is related to superhelical density $\sigma$ as
$\tau=C \omega_0 \sigma$ \cite{endnote2}
and that
B-Z transitions were observed at $\sigma=\sigma_c\sim-0.01$
in the experiment \cite{BZtrans},
we estimate that
$C=\tau_c/\omega_0\sigma_c \sim 130\times 10^{-20}$J.
Then, we obtain
\begin{equation}
K_\text{eq}=\frac{t_\text{Z}}{t_\text{B}}e^{-4 \beta \gamma C \omega_0
 (\sigma-\sigma_c)},
\end{equation}
where 
the coefficient of $\sigma$ in the exponent
is $4 \beta \gamma C \omega_0 \sim 4.5\times 10^2$.
Here we used 
$T= 40^\circ$C and $\gamma=0.37$ [fitted value in Fig.~\ref{slope_t}(d)].
Thus,
the value of the coefficient of $\sigma$ estimated here
qualitatively agrees with
or, more properly speaking,
 is approximately one-fifth of 
the experimental value ($\sim 1.4\times10^3$)
\cite{endnote3}.

In summary,
we have shown
that our nucleation model not only provides 
a clear picture 
for the mechanically induced structural transition in DNA as
the property of a
one dimensional chiral material,
but also offers an explanation for recent experimental observations.

\section{ Summary}
We have constructed the single DNA mechanical model,
which describes the interplay between intrinsic base-pair structures and
global conformations.
With this model,
the mechanical responses of linear DNA to external torques
were simulated and
the singular response near $\tau=\tau_c$
was found
 in the low-temperature region. 
This singularity 
gives rise to instability of the B-Z transition
with minute negative superhelicity, as in the experiment~\cite{BZtrans},
which was explained in terms of
the cooperative effect depending on the average size of the metastable domain.
Furthermore,
the twist dependency of the transition rate was qualitatively explained by
the nucleation picture.
These observations are analogous with the instability of 
a one-dimensional magnetic system:
In low temperature limit,
there occurs an instability at zero external magnetic field,
which, in our system,
 corresponds to the external torque of  $ \tau = \tau_c $ \cite{oneD}. 
From these estimates, we have a clear picture 
for the mechanically induced structural transition in DNA as
one dimensional chiral material.

We expect that 
this model will provide an additional basis for elucidating
the structure-configuration interplay in higher-order biomolecular architectures, 
such as nucleosomes.


\begin{thebibliography}{99}
\bibitem{endnote}
If $B, C$ depend on the structural order parameters, the gauge coupling
and thus ${\cal F}_{s+\rho}$ becomes nonlinear and highly nontrivial.
\bibitem{kinknote}
In the continuum limit,
the kink(antikink) solution for the symmetric potential is given by
$\rho(s) =\pm \omega_0 \tanh(\omega_0^{-1}\sqrt{2V_0/d_1}s)$.
The corresponding domain wall energy  is estimated as
$E_\text{BZ}=\frac{8}{3}\sqrt{2 d_1 V_0}\omega_0 \simeq 0.6 V_0$. 
This value is of the same order as the fitted value in Fig.~\ref{slope_t}.
\bibitem{endnote2}
The equation $\tau=C \rho \sigma$
is derived from
$d\chi/ds=\rho+\tau/C$,
$\sigma=(d\chi/ds-\rho)/\rho$,
and 
$\rho \sim \omega_0$.
\bibitem{endnote3}
The experimental value is obtained by 
finding the exponential curve fitting for
the experiment of $K_\text{eq}$ shown in Fig.~6(B) of Ref.~\cite{BZtrans}.
The experiment was done at temperature of 40$^\circ$C.
\end{thebibliography}
\end{document}